\documentstyle[aps,epsfig,multicol,amsmath]{revtex}

\begin{document}

\date{May 16, 2000}
\draft

\title{Role of Coulomb correlations in the optical spectra of
       semiconductor quantum dots}

\author{\underline{Ulrich Hohenester}~$^{(a)}$ and Elisa Molinari}

\address{
Istituto Nazionale per la Fisica della Materia (INFM) and
Dipartimento di Fisica\\
Universit\`a di Modena e Reggio Emilia, 
Via Campi 213/A, 41100 Modena, Italy\\
$^{(a)}$E-mail: hohenester@unimo.it;  FAX: +39 059 367488}

\maketitle

\begin{abstract} 

We present a consistent theoretical description of few-particle effects
in the optical spectra of semiconductor quantum dots, based on a
direct-diagonalization approach. We show that, because of the strong
Coulomb interaction among electrons and holes, each configuration of
the confined few-particle system leads to its characteristic signature
in the optical spectra. We discuss quantitative predictions and
comparison with experiments for both absorption and luminescence.

\ \\



\end{abstract}


\begin{multicols}{2}
\narrowtext


The strong three-dimensional quantum confinement in semiconductor
quantum dots (QDs) results in a discrete, atomic-like carrier density
of states. In turn, {\em (i)}\/ the coupling to the solid-state
environment (e.g., phonons) is strongly suppressed
\cite{hawrylak:98,bimberg:98} and {\em (ii)}\/ Coulomb correlations
among charge carriers are strongly enhanced.  Indeed, in the optical
spectra of single dots spectrally narrow emission peaks have been
observed (indicating a small environment coupling), which undergo
discrete energy shifts when more carriers are added to the dot
(indicating energy renormalizations due to additional Coulomb
interactions) \cite{few-particle}.

In this paper we discuss how these spectral changes result from
few-particle interactions. A detailed discussion of excitonic and
biexcitonic features in the absorption spectra of parabolic QDs is
presented; luminescence spectra of multi-excitons and multi-charged
excitons are presented, which are compared with experimental data.


The initial ingredients of our calculations are the single-particle
states $\phi_\mu^{e,h}$ and energies $\epsilon_\mu^{e,h}$ for electrons
($e$) and holes ($h$), which we obtain by numerically solving the 3D
single-particle Schr\"odinger equation within the envelope-function and
effective-mass approximations for arbitrary confinement potentials
\cite{rossi:96}.  Next, the few-particle Hamiltonian (containing all
possible $e$-$e$, $e$-$h$, and $h$-$h$ Coulomb matrix elements) is
expanded within the basis of the $\sim$10--20 energetically-lowest
single-particle states, and the few-particle states are obtained by
direct diagonalization of the Hamiltonian matrix (see Appendix). For
simplicity, in the calculation of the few-particle $e$-$h$ states
interaction processes with the dot environment are neglected, and only
a small broadening of the emission peaks is introduced in the
calculation of the optical spectra.

{\em Single excitons.}\/ First, we consider the linear optical
response. Here, a single electron-hole pair (exciton) is created by an
external light field (e.g., laser), which propagates in presence of of
the dot confinement and of mutual Coulomb interactions. The exciton
energies $E_x$ and wavefunctions $\Psi_{\mu;\nu}^x$ are obtained from
the two-particle Schr\"odinger equation:

\begin{equation}\label{eq:exciton}
  (\epsilon_\mu^e+\epsilon_\nu^h)\Psi_{\mu;\nu}^x+
  \sum_{\mu',\nu'}V_{\mu\mu',\nu\nu'}^{eh}\Psi_{\mu';\nu'}^x=
  E_x \Psi_{\mu';\nu'}^x,
\end{equation}

\noindent with the $e$-$h$ Coulomb elements $V^{eh}$ defined in
Eq.~(\ref{eq:coulomb}) of the Appendix. The optical absorption spectra
are then obtained according to Ref.~\cite{rossi:96} from
$\alpha(\omega)\propto\sum_x |M_x|^2{\cal D}_\gamma(\omega-E_x)$, where
$M_x=\sum_{\mu,\nu} \Psi_{\mu;\nu}^x M_{\nu\mu}^{he}$, $M^{he}$ are the
optical dipole elements (see Appendix), and ${\cal
D}_\gamma(\omega)=2\gamma/(\omega^2 +\gamma^2)$ with a phenomenological
damping constant $\gamma$ accounting for interactions with the dot
environment.

Figure 1{\em (a)}\/ shows the linear optical absorption for a
prototypical dot confinement which is parabolic in the $(x,y)$-plane
and box-like along $z$ \cite{dot}; such confinement potentials have
been demonstrated to be a particularly good approximation for various
kinds of self-assembled dots \cite{hawrylak:98,rinaldi:96}. We observe
a series of pronounced absorption peaks ($X_0$, $X_1$, \dots) with an
energy splitting of the order of the confinement energy; an inspection
of the exciton wavefunctions $\Psi_{\mu;\nu}^x$ reveals that the
dominant contribution of excitons $X_0$, $X_1$, and $X_2$ is from the
electron and hole single-particle states $1s$, $1p$, and $(2s,1d)$ (see
inset of Fig.~1{\em (a)}). In analogy with semiconductor quantum wires
\cite{rossi:96}, because of Coulomb interactions the energy separation
between the groundstate exciton $X_0$ and $X_1$ is increased, and
oscillator strength is transferred from peaks of higher energy to those
of lower energy (note that in a pure single-particle picture the ratio
of peak heights follows the degeneracy of the respective shells, i.e.,
$1:2:3:4$); finally, for a discussion of the additional Coulomb-induced
peaks ($X_0^*$, $X_1^*$, \dots) the reader is referred to
Ref.~\cite{hohenester:99b,hohenester:00}.

{\em Biexcitons.}\/ If the dot is populated by two electron-hole pairs,
the carrier states available strongly depend on the $e$-$h$ spin
orientations ($\sigma_\uparrow$, $\sigma_\downarrow$). In the following
we only discuss the case of two electrons (holes) with antiparallel
spin orientations (for parallel spins see
Refs.~\cite{hawrylak:99,hohenester:00}). Then, the biexciton energies
$\bar E_\lambda$ and wavefunctions $\bar\Psi^\lambda$ are obtained from
the $2e$-$2h$ Schr\"odinger equation (accounting for the various
$e$-$e$, $e$-$h$, and $h$-$h$ Coulomb interactions) which, for
conceptual clarity, we write in the exciton basis $x$
\cite{hohenester:00,axt:98}:

\begin{equation}\label{eq:biexcitons}
  (E_x+E_{x'})\bar\Psi_{xx'}^\lambda+
  \sum_{\bar x\bar x'}\bar V_{xx',\bar x\bar x'}
  \bar\Psi_{\bar x\bar x'}^\lambda=
  \bar E_\lambda\bar\Psi_{xx'}^\lambda,
\end{equation}

\noindent with $\bar V_{xx',\bar x\bar x'}$ the exciton-exciton Coulomb
elements \cite{hohenester:00,axt:98}, and $x$, $\bar x$ ($x'$, $\bar
x'$) labeling exciton states with $\sigma_\uparrow$
($\sigma_\downarrow$). Apparently, the exciton-exciton interaction
$\bar V$ in Eq.~(\ref{eq:biexcitons}) is responsible for the
renormalization of the biexciton spectrum. Roughly speaking, the
leading contributions to $\bar V$ are of dipole-dipole character, with
the dipole elements $\mu_{xx'}$ according to the excitonic transitions
from $x$ to $x'$ ($\bar x$ to $\bar x'$) \cite{hohenester:00}; thus, in
general both optically allowed and forbidden (due to wavefunction
symmetry; see also Fig.~1{\em (b)}) excitons with their small and large
values of $\mu$, respectively, contribute to $\bar\Psi^\lambda$.

Figure 2{\em (b)}\/ (2{\em (c)}) shows optical absorption for a dot
which is initially prepared in the $X_0$ ($X_1$) single-exciton state;
this scenario of optically probing a non-equilibrium dot is similar to
the nonlinear coherent optical response, where a strong pump pulse
creates an exciton population at energy $\omega_p$ and a weak probe
pulse monitors the spectral changes due to the induced exciton
population \cite{bonadeo:98}. For the dot initially prepared in state
$X_0$ (Fig.~2{\em (b)}), we observe: At energy $E_{X_0}$ negative
absorption (i.e., gain) due to the removal of the initial exciton
population (i.e., stimulated emission via $X_0+h\nu\to 2h\nu$); the
appearance of an absorption peak $B_0$ on the low-energy side of $X_0$
and of a peak multiplet (labeled $B_1$) at spectral position $X_1$,
attributed to the photo-induced formation of biexcitonic states via
$X_0+h\nu\to B$. To a good approximation, the biexciton groundstate
$B_0$ consists of two groundstate excitons $X_0$ with antiparallel spin
orientations (because of the small value of $\mu$ for optically allowed
excitons the biexciton binding is relatively small); the biexciton
states $B_1$ consist of $e$-$h$ pairs in the $1s$ and $1p$-shells (see
inset of Fig.~1{\em(a)}), where the strong mixing with optically
forbidden excitons (large $\mu$'s) leads to large renormalizations and
to a strong decrease of the oscillator strengths of the absorption
peaks.

{\em Multi-excitons.}\/ Next, we turn to the case of a dot populated by
a larger number of electron-hole pairs. In a typical single-dot
experiment \cite{few-particle}, a pump pulse creates $e$-$h$ pairs in
continuum states (e.g., wetting layer) in the vicinity of the QD, and
some of the carriers become captured in the QD; from experiment it is
known that there is a fast subsequent carrier relaxation due to
environment coupling to the $e$-$h$ states of lowest energy
\cite{bimberg:98}; finally, electrons and holes in the dot recombine by
emitting a photon. By varying in a steady-state experiment the pump
intensity and by monitoring luminescence from the dot states, one thus
obtains information about the few-particle carrier states. From a
theoretical point of view, luminescence involves a process where one
$e$-$h$ pair is removed from the interacting many-particle system and
one photon is created. Thus, luminescence spectra provide information
about $e$-$h$ excitations, in contrast to transport measurements of QDs
\cite{addition-spectra} which only provide information about the
few-particle groundstate.

Fig.~3{\em (a)}\/ shows luminescence spectra for different numbers of
$e$-$h$ pairs (with dot parameters of \cite{dot}; luminescence
intensity computed according to Ref.~\cite{hawrylak:98} and using the
few-particle states of Eq.~(\ref{eq:few-particle})). We assume that
before photon emission the interacting $e$-$h$ system is in its
respective groundstate, i.e., for one $e$-$h$ pair the exciton
groundstate $X_0$; for two $e$-$h$ pairs the state $B_0$; for three
$e$-$h$ pairs approximately a filled $1s$-shell and one $e$-$h$ pair in
the $1p$-shell, etc.; thus, for one $e$-$h$ pair luminescence solely
originates from the decay of $X_0$; for two $e$-$h$ pairs the biexciton
$B_0$ decays into $X_0$, where the emission peak is slightly
red-shifted because of the biexciton binding. In case of three $e$-$h$
pairs the situation is more involved: For recombination of an $e$-$h$
pair in the $1p$-shell, the corresponding luminescence peak is
red-shifted by $\approx$10 meV with respect to $X_1$ because of
exchange corrections of the groundstate energy; for recombination in
the $1s$ shell, after photon emission the dot is in an excited
biexciton state; consequently, the peak multiplet in the luminescence
spectra is determined by the rather complicated density of states of
biexcitonic resonances (see discussion above). Finally, for an
increasing number of $e$-$h$ pairs we observe emission from the $1s$
and $1p$ shells, where the peak multiplet from the $1s$-shell emission
exhibits strong spectral changes as a function of the number of $e$-$h$
pairs. We note that our findings are similar to those obtained in the
strong-confinement limit \cite{hawrylak:99} (the difference for the
$6e$-$6h$ decay is due to the coupling to higher shells, which are
considered in our calculations). Elsewhere \cite{rinaldi:00}, it will
be shown that our calculated luminescence spectra are in good agreement
with experimental single-dot data, with the dots of
Ref.~\cite{rinaldi:96}.

{\em Multi-charged excitons.}\/ We finally discuss luminescence spectra
of multiple-charged excitons. Experimental realization of such carrier
complexes can be found, e.g., in Ref.~\cite{hartmann:00}, where
GaAs/AlGaAs quantum dots are remote-doped with electrons from donors
located in the vicinity of the dot. Employing the mechanism of
photo-depletion of the QD together with the slow hopping transport of
impurity-bound electrons back to the QD, it is possible to efficiently
control the number of surplus electrons in the QD from one to
approximately six \cite{hartmann:00}. Fig.~3{\em (b)} shows
luminescence spectra of charged excitons for a varying number of
surplus electrons and for the prototypical dot confinement of
\cite{dot}. Quite generally, the spectral changes with increasing
doping are similar to those presented for multi-exciton states: With
increasing doping the main peaks red-shift because of exchange and
correlation effects. As in the case of multi-exciton states, each
few-particle state leads to a specific fingerprint in the optical
response. This unique assignment of peaks or peak multiplets to given
few-particle configurations allows to unambiguously determine the
detailed few-particle configuration of carriers in QDs in optical
experiments; this fact is used in Ref.~\cite{hartmann:00} to study the
impurity-dot transport.

{\em Conclusion.}\/ We have presented a detailed study of excitonic and
biexcitonic features in the optical spectra of a parabolic quantum dot.
Luminescence spectra of multi-exciton and multi-charged exciton states
have been analyzed, and we have shown that each few-particle
configuration leads to its specific fingerprint in the optical
response.


{\em Acknowledgements.}\/ This work was supported in part by the EU
under the TMR Network "Ultrafast" and the IST Project SQID, and by INFM
through grant PRA-SSQI. U.H.  acknowledges support by the EC through a
TMR Marie Curie Grant.

\section*{Appendix}

%
%

{\it 1. Matrix elements.}\/ With $\phi^{e,h}$ the single-particle
states for electrons and holes, the optical matrix elements are of the
form \cite{rossi:96} $M_{\nu\mu}^{he}=\mu_o\int d{\bf
r}\;\phi_\nu^h({\bf r})\phi_\mu^e({\bf r})$, with $\mu_o$ the dipole
matrix element of the bulk semiconductor. The Coulomb matrix elements
read:

\begin{equation}\label{eq:coulomb}
  V_{\mu'\mu,\nu'\nu}^{ij}=q_iq_j\int d{\bf r}d{\bf r}'\;
  \frac{{\phi_{\mu'}^i}^*({\bf r}){\phi_{\nu'}^j}^*({\bf r}')
  \phi_{\nu}^j({\bf r}')\phi_{\mu}^i({\bf r})}{\kappa_o|{\bf r}-{\bf r}'|},
\end{equation}

\noindent with $\kappa_o$ the static dielectric constant of the
semiconductor, $i,j=e,h$ and $q_{e,h}=\mp 1$ (note that we have
neglected electron-hole exchange interactions).

{\it 2. Few-particle states.}\/ We compute the few-particle
electron-hole states within a direct-diagonalization approach. With the
creation operators $c^\dagger$ and $d^\dagger$ for electrons and holes,
respectively, we define the $N_e$-electron and $N_h$-hole states
$|\vec\mu\rangle_{N_e}=c_{\mu_1}^\dagger c_{\mu_2}^\dagger\dots
c_{\mu_{N_e}}^\dagger|\Phi_o\rangle$ and $|\vec\nu\rangle_{N_h}=
d_{\nu_1}^\dagger d_{\nu_2}^\dagger\dots d_{\mu_{N_h}}^\dagger
|\Phi_o\rangle$ (vacuum state $|\Phi_o\rangle$; spin degrees of freedom
have not been indicated explicitly), and we keep the $\sim$100
few-particle states of lowest single-particle energies. Next, the
few-particle Hamiltonian ${\cal H}$, accounting for all possible
electron-electron, electron-hole, and hole-hole Coulomb matrix
elements, is expanded within these bases; the few-particle energies
$E_\ell$ and wavefunctions $\Psi_{\vec\mu\vec\nu}^\ell$ are then
obtained through direct diagonalization of the Hamiltonian matrix:

\begin{equation}\label{eq:few-particle}
  E_\ell\Psi_{\vec\mu;\vec\nu}^\ell=
  \sum_{\vec\mu',\vec\nu'}\;_{N_e;N_h}
  \langle\vec\mu;\vec\nu|\;{\cal H}\;
  |\vec\mu';\vec\nu'\rangle_{N_e;N_h}\;\Psi_{\vec\mu';\vec\nu'}^\ell.
\end{equation}


\newpage
\end{multicols}

\begin{figure}
\begin{center}
\end{center}
\caption{
{\em (a)}\/ Linear absorption spectrum for dot [\ref{ref:dot}]
(computed for a basis of 20 electron and 40 hole single-particle
states); the inset shows the single-particle wavefunctions of lowest
energy (states $1p$ are double degenerate, states $1d$ and $2s$ are
triple degenerate). {\em (b)}\/ Contour plot of the exciton
wavefunctions $\sum_x|\Psi^x({\bf r},{\bf r})|^2_{ {\bf r}=(X,0,0)}
{\cal D}_\gamma(\omega-E_x)$; because of symmetry only a small portion
of the excitons couples to the light [\ref{ref:hohenester:00}].
}
\end{figure}

\begin{figure}
\begin{center}
\end{center}
\caption{
Optical absorption spectra for QD initially prepared in the
{\em(a)}\/ vacuum state (i.e., linear absorption), {\em(b)}\/ exciton
grondstate $X_0$, {\em(c)}\/ state $X_1$ (i.e., nonlinear absorption).
For a discussion see text.
}
\end{figure}


\begin{figure}
\caption{ 
Luminescence spectra for QD and for different {\em(a)}\/ multi-exciton
and {\em(b)}\/ multi-charged excitons states. We assume that before
photon emission the electron-hole system is in its groundtstate
(i.e., {\em(a)}:\/ 
$(1e_\uparrow;1h_\downarrow)$, 
$(1e_\uparrow,1e_\downarrow;1h_\downarrow,1h_\uparrow)$,
$(2e_\uparrow,1e_\downarrow;2h_\downarrow,1h_\uparrow)$,
$(3e_\uparrow,1e_\downarrow;3h_\downarrow,1h_\uparrow)$,
$(3e_\uparrow,2e_\downarrow;3h_\downarrow,2h_\uparrow)$, and
$(3e_\uparrow,3e_\downarrow;3h_\downarrow,3h_\uparrow)$;
{\em(b)}:\/ 
$(1e_\uparrow;1h_{\uparrow,\downarrow})$,
$(1e_\uparrow,1e_\downarrow;1h_{\uparrow,\downarrow})$,
$(2e_\uparrow,1e_\downarrow;1h_{\uparrow,\downarrow})$,
$(3e_\uparrow,1e_\downarrow;1h_{\uparrow,\downarrow})$,
$(3e_\uparrow,2e_\downarrow;1h_{\uparrow,\downarrow})$,
$(3e_\uparrow,3e_\downarrow;1h_{\uparrow,\downarrow})$)
}
\end{figure}


\begin{thebibliography}{10}
\bibitem{dot} The confinement energies due to the in-plane parabolic
potential are $\omega_o^{e}=20$ meV for electrons and
$\omega_o^{h}=3.5$ meV for holes (with this choice, electron and hole
wavefunctions have the same lateral extension; we use material
parameters for GaAs), and the quantum-well confinement along $z$ is
such that the intersubband splittings are much larger than
$\omega_o^{e,h}$. \label{ref:dot}
\end{thebibliography}
\end{document}